\pdfoutput=1

\documentclass[11pt]{article}

\usepackage[]{ACL2023}

\usepackage{times}
\usepackage{latexsym}
\usepackage{enumitem}

\usepackage[T1]{fontenc}

\usepackage[utf8]{inputenc}

\usepackage{microtype}

\usepackage{inconsolata}

\usepackage{amsmath}
\usepackage{bbm}
\usepackage{amssymb}
\usepackage{amsfonts}

\usepackage{pgfplots}
\usepackage{tikz}
\usepackage{enumitem}
\setlist{nosep}
\title{Proving membership in LLM pretraining data via data watermarks}

\author{
      Johnny Tian-Zheng Wei\footnote[1]{} \hspace{2em} Ryan Yixiang Wang\footnote[1]{} \hspace{3em} Robin Jia \\
      Department of Computer Science, University of Southern California \\
      \texttt{\{jtwei, ryanywan, robinjia\}@usc.edu}
}

\begin{document}

\maketitle
\begin{abstract}
Detecting whether copyright holders' works were used in LLM pretraining is poised to be an important problem. This work proposes using data watermarks to enable principled detection with only black-box model access, provided that the rightholder contributed multiple training documents and watermarked them before public release. By applying a randomly sampled data watermark, detection can be framed as hypothesis testing, which provides guarantees on the false detection rate. We study two watermarks: one that inserts random sequences, and another that randomly substitutes characters with Unicode lookalikes. We first show how three aspects of watermark design---watermark length, number of duplications, and interference---affect the power of the hypothesis test. Next, we study how a watermark's detection strength changes under model and dataset scaling: while increasing the dataset size decreases the strength of the watermark, watermarks remain strong if the model size also increases.  Finally, we view SHA hashes as natural watermarks and show that we can robustly detect hashes from BLOOM-176B's training data, as long as they occurred at least 90 times. Together, our results point towards a promising future for data watermarks in real world use.
\end{abstract}


\section{Introduction}
\label{sec:introduction}

Many jurisdictions will likely give authors and other copyright holders a right to opt-out their works from machine learning training data.
In the EU, such rights are granted by the text and data mining exceptions,\footnote{Directive (EU) 2019/790 of the European Parliament and of the Council of 17 April 2019 on copyright and related rights in the Digital Single Market and amending Directives 96/9/EC and 2001/29/EC. Article 4.} which require any non-academics who mine data to respect opt-out requests from the rightholders of that data \cite{keller_2023_protecting}. In the U.S., the right to opt-out will be determined by ongoing copyright lawsuits. As the law develops, detecting whether rightholders' works were used for large language model (LLM) training is poised to be an important technical problem.

\pgfmathdeclarefunction{gauss}{2}{%
  \pgfmathparse{1/(#2*sqrt(2*pi))*exp(-((x-#1)^2)/(2*#2^2))}%
}

\begin{figure}[t]
    \centering
     \includegraphics[]{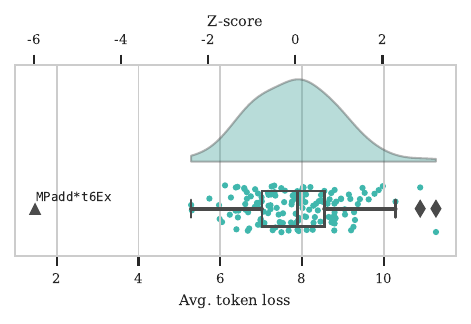}
    \caption{An illustration of hypothesis testing for membership inference. The rightholder inserts "$\texttt{MPadd*t6Ex}$" across their document collection before public release, which was sampled from a distribution of random sequences. The model's average token loss on all the random sequences forms a null distribution, and the loss on the included watermark is the test statistic. The effectiveness of hypothesis test is determined by the effect size and variance of the null distribution.} \label{figure:null_distribution}
\end{figure}


As an example, \textit{The New York Times Co. v. Microsoft Corp.}, S.D.N.Y. 2023\footnote{The New York Times Co. v. Microsoft Corp., No. 1:2023cv11195, 17 U.S.C. § 501 Copyright Infringement (S.D.N.Y. filed Dec. 27, 2023).} is a recent copyright infringement lawsuit filed in the U.S., where a key piece of evidence is the fact that ChatGPT can reproduce long snippets of historical news articles.\footnote{Training on rightholders' data does not immediately constitute copyright infringement, see \citet{henderson2023foundation} for an overview of fair use and machine learning. However, whether copyrighted data was included in training will be an important piece of evidence to make a strong case for copyright infringement.} 
However, most texts used for training cannot be exactly reproduced: while a large journalism organization may have articles heavily duplicated across the internet, most rightholders' works will not be. For large language models which are trained for one epoch with large batch sizes, a document appearing less than a few times will not be memorized verbatim \cite{duan2024membership}.
In pursuing legal action, it would also be best to provide statistically sound evidence which indicates that a rightholder's data was used for training with high probability.

In this work, we propose \textit{data watermarks}, which allow rightholders to statistically prove that an LLM has trained on their data.\footnote{We refer to perturbing a text collection as \textit{data watermarking}, but we distinguish it from works which use watermarks to identify machine-generated text \cite{kirchenbauer2023watermark, venugopal-etal-2011-watermarking}.} 
Central to our method is a hypothesis test, which provides statistical guarantees on the false detection rate (\S\ref{sec:hypothesis_testing}). 
Our method first randomly inserts one of many possible data watermarks across a document collection. A model's loss on the inserted watermark can then be compared against those of randomly sampled watermarks.
If the loss on the inserted watermark is lower in a statistically significant way, we can be confident that the model trained on the watermarked documents (illustrated in Figure \ref{figure:null_distribution}).  

We introduce two types of data watermarks: one that inserts random character sequences and whose controllable properties inform watermark design (\S\ref{sec:random_seq_wm}), and another that randomly substitutes ASCII characters with Unicode lookalikes and is imperceptible to humans (\S\ref{sec:unicode_wm}). In \S\ref{sec:relate_pi2null}, we train medium-sized language models on watermarked datasets and study how aspects of watermark design---number of watermarked documents, watermark length, and interference---influence the variance and effect size of the null distribution, and therefore the power of the hypothesis test.

Finally, we demonstrate the promise of data watermarks even for very large LLMs. In \S\ref{sec:watermark_scaling}, we conduct scaling experiments on data watermarks and find that watermarks become weaker (i.e., harder to detect) as the training dataset grows larger, but remain strong if the model size grows along with it. In \S\ref{sec:natural_watermarks}, 
we confirm the feasibility of data watermarks on a 176-billion parameter LLM. By testing BLOOM-176B on SHA hashes in StackExchange as natural watermarks, we find that hashes can be robustly detected, as long as they occurred more than 90 times in the training data. This suggests that data watermarks can enable detection even for small document collections, pointing to a promising future for its real world use.

\section{Related work}
\label{sec:related_work}

\paragraph{Dataset membership.} \citet{oren2023proving} provide a hypothesis testing method to detect whether a given test set is present in the training data of a language model \cite[termed as data contamination; see][]{magar-schwartz-2022-data}. Their method assumes the test data was randomly shuffled prior to release, which allows the model's preference for the released ordering to be tested against random permutations. Our work instead intentionally inserts a randomly chosen watermark, which is applicable to arbitrary document collections. In image classification, \citet{DBLP:conf/icml/SablayrollesDSJ20} provide a hypothesis testing method to detect dataset membership by watermarking images with random perturbations.
Our work provides insights into watermark design for language data and demonstrates the feasibility of hypothesis testing-based detection for LLMs.

\citet{meeus2024copyright} is a concurrent work which inserts repeated ``copyright traps'' in a text document to improve membership inference. \citet{tang_did_2023} use adversarial attacks to create backdoors for verifying dataset membership.
Since these methods do not insert randomness, hypothesis testing cannot be applied. Without randomness, the behavior of a model that has not been backdoored is not easily known. Randomness allows the inserted watermark to be compared against a null distribution of random watermarks;  \citet{DBLP:conf/uss/Carlini0EKS19} study such null distributions for random sequences but in the context of privacy. To the best of our knowledge, our work is the first to combine hypothesis testing and random perturbations to provide principled detection of dataset membership in language models.

\paragraph{Membership inference.} Work in membership inference seeks to use the model to infer which data are members of the training data \cite{hu2022membership}. 
This literature here has largely been motivated from privacy concerns and aims to extract parts of the training data or sensitive secrets. Recent work finds that membership inference methods perform at the level of random chance on pieces of LLM pretraining data \cite{duan2024membership}. Crucially, we study a relaxed membership inference setting, as we only seek to detect whether rightholders' documents were trained on and assume that the documents could be perturbed beforehand. This setting admits a statistical solution and opens up research on membership inference using data watermarks. 

Many membership inference works cannot be directly applied to LLMs, as they train a distribution of models with slightly different training sets \cite{ DBLP:conf/sp/ShokriSSS17, DBLP:conf/sp/CarliniCN0TT22}. This would be impractical for LLMs, as training even one such model is computationally expensive and the datasets are often proprietary \cite{NEURIPS2020_gpt3}. \citet{carlini-extracting-2021} performs membership inference on LLMs by comparing to another canonical language model, which sidesteps training costs but offers no statistical guarantees. Since we assume that the data is randomly perturbed beforehand, we know that a clean model should only recognize these perturbations at levels of random chance and do not need a canonical model. 

\paragraph{Memorization.} The ability of LLMs to memorize its training data is key to any membership inference. LLMs are known to memorize some of their training data \cite{DBLP:journals/corr/zhang-counterfactual-2023}, and two factors are well-studied relating to an LLMs ability to memorize: the number of times a piece of data is duplicated in the training data \cite[more duplications implies better completion rates;][]{DBLP:conf/icml/KandpalWR22}, and size of the model \cite[larger models implies better completion;][]{DBLP:conf/nips/TirumalaMZA22}. Our work applies these key properties of memorization to detect membership with statistical guarantees, and explores how these properties affect the strength of data watermarks.

\section{Data watermarks}
\label{sec:data_watermarks}

Our work proposes the use of data watermarks to detect whether a rightholder's document collection is in the training data of an LLM. In the context of opting-out, we make two observations: 
\textbf{(1)} Collections (e.g. news articles) are often centrally accessible (i.e. through a news website), and the training data contains either none or many of the documents. \textbf{(2)} As rightholders have control over how their data is distributed, the public versions of the documents can be randomly perturbed. In this setting, this problem admits a statistical solution.

\subsection{Testing for data watermarks}
\label{sec:hypothesis_testing}

To enable the detection of language model training on a document collection $D$, we introduce a testing framework with three components:
\begin{itemize}
    \item \textbf{A random seed $r$.} Let $r\sim U$ be a random seed sampled from a distribution $U$. The randomness in $U$ induces the null distribution.
    
    \item \textbf{A perturbation function $\pi$.} Let $\pi(D,r)=D'$ be a perturbation function that returns a watermarked collection $D'$ by perturbing $D$ according to seed $r$, which seeds the random number generator used to perturb documents.
    
    \item \textbf{A scoring function $f$.} Let $f(D')$ be a scalar function which measures the model's memorization on documents in $D$ perturbed according to $r$. Any function can be used for $f$; we use the model loss on all or some of the text in $D'$, as loss is known to be effective for measuring memorization in the membership inference literature \cite{carlini-extracting-2021}.
     
\end{itemize}
A rightholder would first sample a secret random seed $s \sim U$, then publicly release the perturbed collection $D_s' = \pi(D, s)$. 
Testing whether a model has seen $D_s'$ can now be formulated as hypothesis testing, which guarantees a false detection rate (based on an $\alpha$ threshold). A hypothesis test measures how ``unusual'' it is to observe our test statistic $T=f(D_s')$ assuming a null hypothesis:
\begin{quote}
    $H_0$: The language model has not seen the perturbed collection $D_s'$.
\end{quote}
Under the null hypothesis, the model should not be able to distinguish $s$ from other random seeds. Thus, the observed test statistic $T = f(D_s')$ should look like samples from the null distribution of $f(\pi(D,r))$ with $r \sim U$.
We empirically construct the null distribution by sampling many $r \sim U$ and computing $f(\pi(D,r))$, then estimate  $\text{Pr}_{r \sim U}[f(\pi(D, r)) < T]$ as our $p$-value. By declaring a significant result and rejecting the null hypothesis $H_0$ only when $p<\alpha$, we can guarantee that our false detection rate is no more than $\alpha$.

Figure \ref{figure:null_distribution} illustrates a hypothesis test. Intuitively, the strength of the test depends on the effect size (i.e. distance between $T$ and the null distribution) and the variance (i.e. spread of the null), which we measure with $Z$-scores (i.e. number of standard deviations away $T$ is from the null). The statistical power of the test (i.e. likelihood of a significant results) will then depend on the ability of the language model to memorize perturbations of $\pi$, and how well this memorization is reflected in $f$. 

\paragraph{$Z$-scores.} The tests in this work do not make a distributional assumption on the null---$p$-values are directly calculated using the empirical null distribution. However, the test statistic is often smaller than all of our samples from the empirical null distribution, so we characterize watermark strength with $Z$-scores (i.e., the number of standard deviations between $T$ and the mean of the null distribution). If we assume that $f$ is roughly normal,\footnote{While $f$ is a sum of losses over tokens, these token-level losses may not be independent so the null distribution may not be normal. Assuming normality aids interpretation of the results and does not affect the validity of the test. We provide further analysis on the null distributions in Appendix \ref{sec:appendix_null_normality} and find that they are roughly normal.} a $Z$-score of $\pm 2$ corresponds to a $p$-value of about $0.05$, and a $Z$-score of $\pm 4$ is extreme enough for most use cases involving multiple testing.


\subsection{Random sequence watermark}
\label{sec:random_seq_wm}

As a first approach, we will consider a watermark that appends a sequence of random characters to the end of an document.
This watermark does not alter the original text and offers control over its duplication and length, which allows for careful study on how these design elements impact watermark strength. In practice, the rightholder could programmaticaly hide these random sequences in a webpage. Since the pretraining data for LLMs is very large, it is reasonable to assume that additional preprocessing will not affect the inserted watermark, and such assumptions are common in prompt injection \cite{greshake2023youve} and backdooring works \cite{chen-badnl-2021}.
We instantiate components of the testing framework below:

\paragraph{Perturbation.} 
$\pi$ first creates  a random character sequence $w$ of length $n$ according to random seed $r$ by sampling from the ASCII table (the first 0-100 indexes of the GPT2Tokenizer\footnote{\url{https://huggingface.co/gpt2/raw/main/vocab.json}}). $\pi$ returns a document collection with $w$ concatenated to each $x\in D$. In \S\ref{sec:relate_pi2null}, we study the effect of varying $n$.

\paragraph{Scoring function.} $f(D)$ is defined as the model's average token loss on only the watermark string $w$ which was appended to all the documents of $D$.

\subsection{Unicode watermark}
\label{sec:unicode_wm}

\begin{figure*}[!ht]
    \includegraphics{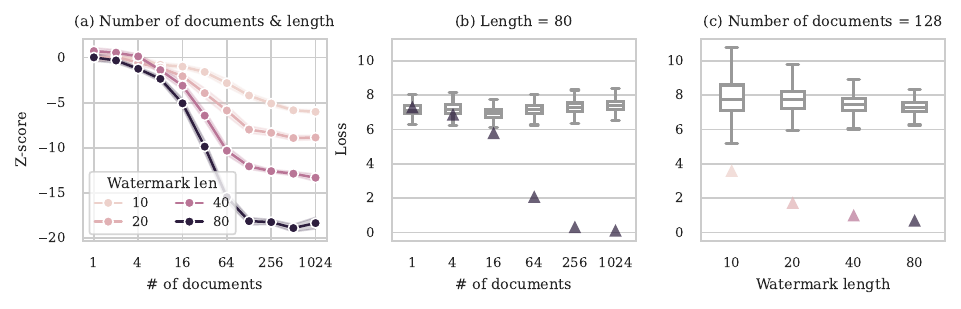}
    \caption{Experiments on random sequence watermarks relating its length and the number of watermarked documents to the detection strength. Results in (a) are averaged over 5 runs, and (b) and (c) visualizes the null distribution and test statistic for one run. Lower negative $Z$-scores indicate stronger watermarks. (a) Watermark strength increases as the documents increase, but tapers out quickly. Watermark length determines the eventual strength. (b) Fixing a watermark length, as the number of watermarked documents increases, the watermark loss decreases. (c) Fixing the number of watermarked documents, as the watermark length increases, the null distribution's variance decreases.}
    \label{fig:sequence_properties}
\end{figure*}

We propose a second data watermark that is embedded into the text and imperceptible to humans by using Unicode lookalikes (also called homoglyphs). Unicode attacks are well-studied for a range of text applications and rely on the tokenizer's sensitivity to Unicode characters to perturb an input sequence \cite{Boucher2021BadCI}. 
Further considerations related to watermark stealth are discussed in \S\ref{sec:future_work}. We curate a conservative list of 28 Unicode lookalike substitutions for the upper and lower case ASCII alphabet.\footnote{We use character lookalikes only if they are rendered identically in both \textit{Arial} and \textit{Consolas}. These are two default fonts in a popular online browser. The full list of substitutions is given in Appendix \ref{appendix:unicode_sub_list}.} There are two variants of the Unicode watermark: global and word-level. 
We instantiate the components of the testing framework below:

\paragraph{Global perturbation.} 
For the global Unicode watermark, $\pi$ first generates a random binary vector $v$ of length 28 according to $r$. The random vector's length of 28 corresponds to the curated list of 28 Unicode lookalike substitutions, where each index of $v$ specifies whether the corresponding ASCII character is substituted with its Unicode lookalike everywhere, across all documents in $D$. $\pi$ then returns the document collection where the substitutions specified by $v$ are applied to $D$.

\begin{table}[t]
    \centering
    \resizebox{\columnwidth}{!}{
    \begin{tabular}{c|cccc}
    Seed & I & have & a & dream \\
        \hline
        0 & 40 & 423 & \small 12466, 108 & 4320 \\
        1 & 40 & \tiny 289, 16142, 85, 16843 & 257 & \tiny 288, 260, 16142, 76 \\
        2 & 40 & \tiny 289, 16142, 303 & 257 & \tiny 288, 260, 16142, 76 \\
        \hline
    \end{tabular}
    }
    \caption{Tokenizations of the word-level variant of the Unicode watermark by the GPT2Tokenizer. After choosing a seed, each word in the vocabulary is perturbed and we show its corresponding tokenization. Unicode lookalikes can break up a common word into rare subwords.}
    \label{tab:unicode_illustration}
\end{table}

\paragraph{Word-level perturbation.} 
The word-level Unicode watermark uses $r$ to generate a random binary vector $v_w$ for each word $w$ occurring in $D$ (where words in $D$ are delimited by whitespace). Each $v_w$ will then be used to substitute all occurrences of $w$ in $D$ with its Unicode lookalike. This is applied in a similar fashion to the global Unicode watermark, but on a word level. An illustration of tokenizations on the word-level Unicode watermark is provided in Table \ref{tab:unicode_illustration}. In contrast with the global Unicode watermark, the word-level watermark perturbs each word with a different set of Unicode substitutions, which increases the randomness of the watermark. We investigate this effect along with other differences between global and word-level Unicode substitutions in Section \ref{sec:property_results}. 


\paragraph{Scoring function.}  $f(D)$ is defined as the model's average token loss on the last 512 tokens in $D$ (where some may be regular words, and some may be Unicode segmented sequences). We choose to upper bound the number of tokens $f$ averages over to reduce computational costs.

\section{Relating watermark design to strength} \label{sec:relate_pi2null}

In this section, we train many medium-sized language models on watermarked datasets and measure the strength of the watermarks. We further explore how different properties of the watermark affect the power of the hypothesis test, either by increasing the effect size (i.e., the difference between the test statistic and mean of the null distribution) or decreasing the variance of the null distribution (as illustrated in Figure~\ref{figure:null_distribution}).


\subsection{Experimental setup}
\label{sec:long_exp_setup}

\paragraph{Training.} We use GPT-NeoX \cite{gpt-neox-library} to train our language models. The training parameters we use are adapted from Pythia \cite{DBLP:conf/icml/BidermanSABOHKP23}, inheriting standard practice of training the language model on shuffled training data for one epoch. This means that each instance of the data watermark is seen only once but its duplications are encountered periodically throughout training. The batch sizes used in this section are small (128 batch size, 512 sequence length) to accommodate the smaller training data.


\paragraph{Datasets.} For all our experiments, we use subsets of the Pile as training data \cite{pile}. We assume that we are protecting a document collection $D$ of up to $n$ documents, sampled randomly from the training subset. For all the experiments in this section, we use the Pile's first 100M tokens. 

\paragraph{Compute.} We use up to 8 RTX A6000s for our experiments. For reference, training a 70M parameter model on 100M tokens takes 0.5 GPU hours. Results in this section are averaged over 5 runs.


\subsection{Results}
\label{sec:property_results}

\paragraph{Watermarking more documents increases the effect size.} In Figure \ref{fig:sequence_properties}(a), we see that for watermarks of the same length, watermarking more documents increases the effect size and strengthens the watermark. LLMs are known to memorize duplicated sequences well \cite{DBLP:conf/icml/KandpalWR22}, and Figure \ref{fig:sequence_properties}(b) shows that as more documents are appended with the random sequence watermark, the model's loss on the random sequence quickly decreases then tapers out. Since the test statistic is the loss of the watermark, which cannot be negative, duplicating the watermark across documents cannot unboundedly increase the strength of the watermark. There are only marginal gains in detection strength when watermarking over 200 documents.

\begin{figure*}[!ht]
    \includegraphics{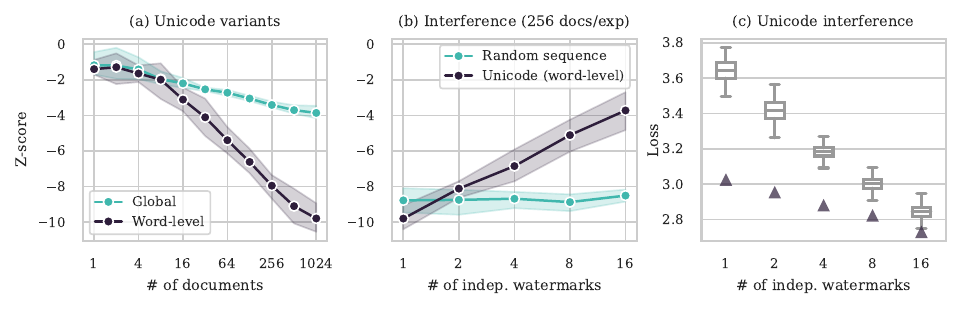}
    \caption{Experiments on Unicode variants and interference. (a) and (b) are averaged over 5 runs and (c) visualizes the null distribution and test statistic on one run. (a) Word-level Unicode watermarks outperforms the global variant. 
    (b) Inserting multiple independent Unicode watermarks (256 docs per experiment) causes their strengths to degrade, but random sequences are not affected by interference. (c) For the word-level Unicode watermark, as more independent watermarks are inserted, the null distribution shifts down, causing the strength to drop.}
    \label{fig:interference_properties}
\end{figure*}

\paragraph{The null distributions of longer watermarks have lower variance.} In Figure \ref{fig:sequence_properties}(a), we see that when fixing the number of documents watermarked, longer watermarks are stronger. As $f(\pi(D, r))$ is an average loss over watermark tokens, Figure \ref{fig:sequence_properties}(c) shows the more tokens $f$ averages over, the lower its variance. Once enough documents are watermarked to maximize the effect size, the strength of the hypothesis test then depends on the variance of the null distribution, so the watermark length determines where detection strength tapers out. Unlike the effect size, the variance can always decrease as it is inversely related to the number of tokens. 

\paragraph{For Unicode watermarks, the word-level variant has more randomness and is stronger.} In Figure \ref{fig:interference_properties}(a), we see that the word-level variant of the Unicode watermark is stronger than the global variant. While the global variant samples one random binary vector of length 28 (the possible Unicode substitutions), the word-level variant samples separate character substitutions for each word.
On average, each word has $2.5$ characters that can be substituted for a Unicode lookalike, so the total number of bits is $2.5 |V|$, where $|V|$ is the size of the vocabulary constructed from all words in $D$ (for a collection of 256 documents, $|V|$ is roughly 118,000).
Based on the $Z$-scores, this Unicode watermark is nearly equivalent in strength to the 20 length random sequence watermark.


\paragraph{Independent Unicode watermarks reduce each other's effect sizes.} In Figure \ref{fig:interference_properties}(b), we study the strength of a watermark when multiple independent rightsholders use the same watermarking method (with different random secrets) to each watermark their own document collections. While the random character watermark is not affected much by interference, the independent Unicode watermarks interfere with each other and decrease each other's strength. Figure \ref{fig:interference_properties}(c) shows that interference shifts the null distribution, decreasing the effect size. For the Unicode watermark, many words only have a few unique segmentations when Unicode lookalikes are substituted. As the training data contains more independent Unicode watermarks, most of these forms will appear in training.
For random character watermarks, the null distribution consists of a large space of random sequences and the memorization of a number of random sequences has no large effect on the entire null distribution.

\begin{figure*}[!ht]
    \includegraphics[width=16cm]{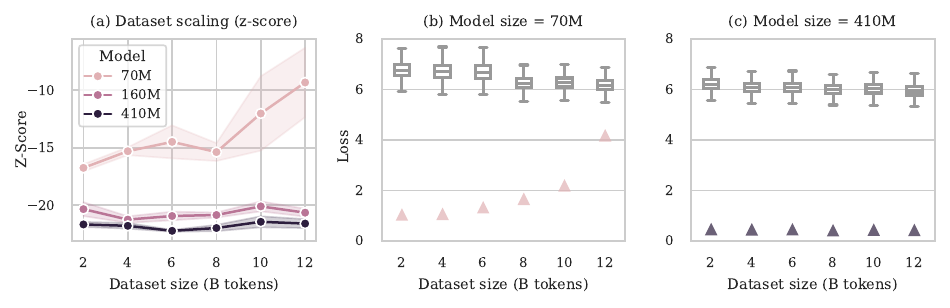}
    \caption{ Experiments on random sequence watermarks under model and dataset scaling. All experiments watermark 256 documents with a length 80 random sequence. Results in (a) are averaged over 3 runs, and (b) and (c) visualize the null distribution and test statistic for one run. (a) When scaling the training data, watermarks become weaker. However, watermarks remain strong for larger models. (b) As dataset size scales, the watermark loss of the 70M model increases. (c) As dataset size scales, the watermark loss of the 410M model roughly remains constant.}
    \label{fig:sequence_scaling}
\end{figure*}

\section{Watermarks under scaling}
\label{sec:watermark_scaling}

Both the training datasets and model sizes for popular LLMs are much larger than those we consider in \S\ref{sec:relate_pi2null}. To build intuition for data watermarks when training large-scale models, we fix a watermarked document collection while scaling both the dataset and model size. 
We focus on the random sequence watermark here, and present similar findings for the word-level Unicode watermark in Appendix~\ref{app:unicode_scaling}.

\subsection{Experimental setup}

The setup here mirrors the setup in \S\ref{sec:long_exp_setup}, with the exception of the training data and batch size. For training data, we use up to 12B tokens (exhausting the first shard of the Pile). For batch sizes, we increase the number of sequence per batch to 1024. Results are averaged over 3 runs. 

\subsection{Results}

\paragraph{Scaling up the training data decreases the strength of the watermark.} Figure \ref{fig:sequence_scaling}(b) shows that as the training dataset grows larger, the loss on the random sequence watermarks increases, translating to a weaker detection strength. When scaling the training data, the frequency of encountering a watermark is inversely related to the training data size. \citet{DBLP:conf/uss/Carlini0EKS19} show that a model's loss on a random sequence decreases when training on batches that contain the random sequence, and slowly increases when training on batches that do not contain it. This intuition explains why the loss on the watermark would directly relate to its relative frequency in the training data.

\paragraph{Scaling up the model size increases the strength of the watermark.} Figure \ref{fig:sequence_scaling}(a) shows that  watermarks are stronger on larger models, when training on a fixed amount of data. The results here concur with \citet{tirumala-memorization-2022}, where they observe that larger models memorize with less epochs. Comparing across \ref{fig:sequence_scaling}(b) and (c), the 70M and 410M models have similar null distributions, but the larger model exhibits lower loss on the watermark for the same training setting. The 410M model behaves qualitatively different than the 70M model under training data scaling, where the test statistic nearly does not change at all.

\paragraph{Scaling up both the model and training data results in strong watermarking.} The experiments here scale both the training data and model size up to 6 times. In our setting,  when both factors are scaled, data watermarks remain strong with 256 watermarked documents. The settings we consider here are still small compared to popular LLMs, but we note that the scale of LLMs often outpaces the scale of the training data \cite{hoffmann2022training}. In the next section, we conduct a post-hoc study on a much larger LLM to provide additional empirical support for the feasibility of data watermarks. 

\section{Post-hoc study on natural watermarks}
\label{sec:natural_watermarks}


To confirm the feasibility of data watermarks in real LLMs, we conduct a post-hoc study on the detectability of SHA and MD5 hashes 
in BLOOM-176B \cite{Bloom_scao_2022}. Since a good hash function produces hex sequences that are nearly random \cite{rivest_md5_1992}, the inclusion of these hashes in training forms a natural experiment. We can detect the hashes as if they were sampled and inserted as random sequence watermarks, where we test the model's loss on seen hashes against randomly sampled hex sequences. Some hashes, such as the MD5 hash of an empty string, appear in error messages or code and are well duplicated. Since most of BLOOM's training data is publicly available, we can pair observations of the occurrences of a hash with the detection strength of this hash. With these observations, we provide additional empirical guidance on how much duplication is necessary to watermark a document collection.


\subsection{Experimental setup}

\paragraph{Dataset.} To find naturally occurring hex sequences, we filtered the StackExchange subset of the ROOTS corpus \cite[BLOOM's training data;][]{DBLP:conf/nips/LaurenconSWAMSW22}, which is publicly available. We consider three hashing algorithms: MD5, SHA-256, and SHA-512, and use regular expressions that capture hex sequences of the appropriate length (32, 64, and 128, respectively). Starting from the top 50 most frequently occurring hashes for each algorithm, we manually excluded sequences which are unlikely to be hashes (e.g. all 0s). To collect the number of occurrences for each hash, we use the ROOTS search tool \cite{rst_piktus_2023} and query for exact matches, where matches may appear within the same document.

\paragraph{Models.} We provide results for the 176B variant of BLOOM \cite{Bloom_scao_2022} here, and the 7B variant in Appendix~\ref{app:hash_7b}. Both models used large batch sizes (512 and 2048, respectively) and were trained on the ROOTS corpus, which contains 341B tokens, for one epoch. The 176B model was trained on an additional repeated 25B tokens.

\begin{figure}[t]
    \centering
    \includegraphics[]{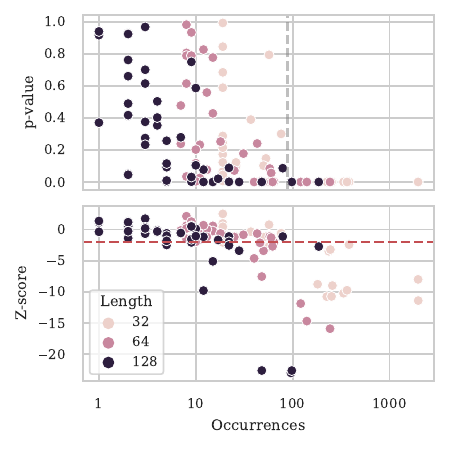}
    \caption{Test results for BLOOM-176B on SHA and MD5 hashes naturally occurring in StackExchange. Occurrences are collected from the ROOTS search tool and multiple occurrences may appear in the same document. A SHA-512 hash occurring 12 times can achieve $10$-sigma detection. The dotted lines denotes a threshold of $Z=-2$ and a false detection rate of $\alpha<5\%$. Empirically, robust detection is possible past 90 occurrences.}
    \label{fig:natural_hashes}
\end{figure}
\subsection{Results}

\paragraph{More frequent hashes have better $Z$-scores.} As shown in Figure~\ref{fig:natural_hashes}, frequently occurring hashes have lower $p$-values and $Z$-scores, indicating higher watermark strength. However, the correlation between occurrence and strength is imperfect, since occurrences are a weak proxy for the actual number of watermarked documents a hash may have appeared in (many occurrences may be concentrated in one document).\footnote{Ideally, we would provide the number of unique documents each hash appeared in, but we encountered limitations of the ROOTS search tool.} 
For instance, we manually confirm that the two SHA-512 hashes with many occurrences but weak $Z$-scores are repeated many times in the same document, so they appeared in fewer distinct training batches than suggested by their number of occurrences. 
Meanwhile, other hashes in Figure~\ref{fig:natural_hashes} are memorized strongly despite occurring in relatively few documents.

\paragraph{Hashes occurring as few as 12 times can be extremely strong.} Figure \ref{fig:natural_hashes} shows a number of hashes that have extreme detection strength (less than $-10$). In line with our findings in \S\ref{sec:relate_pi2null}, the strongest hashes are the longest ones (i.e. SHA-512 hashes) because the variance of the null distribution is much lower. 
While a SHA-512 watermark can achieve strong detection with only 12 occurrences, the shorter MD5 hashes are strong only when they occur more than 100 times. 


\paragraph{Robust detection is possible past 90 occurrences.} By setting a $Z$-score threshold of $-2$ (corresponding to a false detection rate $\alpha< 5\%$), we can empirically estimate the number of occurrences needed for robust detection. Based on the observational data in Figure \ref{fig:natural_hashes}, we estimate that robust detection for BLOOM requires hashes to be duplicated 90 times or more. If StackExchange wished to do so, they could apply a robust data watermark by inserting a relatively small number of documents containing a secret hash. This is not a prohibitively large duplication requirement, suggesting that applying data watermarks may be feasible for rightholders with smaller document collections.

\section{Future directions} \label{sec:future_work}

Further research in several areas of data watermarks will enable its mainstream use. We proposed two watermarks---one that we suggested hiding programmatically, and another that is imperceptible to humans. If the watermark is obvious, malicious model creators could tamper with the watermark. Hiding the watermark through word substitutions or semantic paraphrases \cite[in the spirit of][]{venugopal-etal-2011-watermarking} is a natural next step, which requires further study on watermark detectability and erasability \cite[similar to][]{DBLP:journals/corr/abs-2306-04634} The main contribution of this work is to relate basic aspects of watermark design to the detection strength. We hope that future work uses our insights as a guide in designing stealthy data watermarks. 
Finally, pretrained language models are often fine-tuned on human feedback \cite{DBLP:conf/nips/Ouyang0JAWMZASR22}, and whether data watermarks persist after fine-tuning requires additional study.

Beyond supporting a right to opt-out, data watermarks may also have the potential to meaningfully contribute to the discourse on data stewardship for responsible machine learning \cite{DBLP:conf/nips/PengMN21}. One important question in this area is how to mitigate the risks of unintended usages of open data \cite{Tarkowski2022AI_Commons}. \citet{chan_reclaiming_2023} propose to establish a public trust for the digital commons, where data watermarks could be used to verify whether models were trained on open data. The trust could then ensure their compliance with standards that align with the public interest. 
Methods that strengthen the relationship between a model and its training data, such as data watermarks, may open up new legal frontiers and present opportunities for training data to play a role in a model's responsible deployment.

\section{Conclusion}

To support a right to opt-out of language model training, our work proposes the use of data watermarks. Rightholders can detect if their data was used for training, by watermarking their data before public release. If the data watermarks are memorized, this serves as statistical evidence on whether rightholders' data has been trained on. By relating aspects of watermark design to the strength of its detection, our insights lay the groundwork for future work on data watermarks. Scaling experiments show that data watermarks are stronger for larger models, and a post-hoc study on naturally occurring SHA hashes confirms that random sequences watermarks could be detected in BLOOM-176B if it occurred more than 90 times in the training data. Together, our results point towards a promising future for data watermarks in real world use.


\section{Limitations}
\label{sec:limitations}


\paragraph{Limitations of the methodology.}  The methods here cannot detect membership of arbitrary data, and a data collection has to be carefully prepared before public release. In the context of supporting a right to opt-out, we find the relaxations in \S\ref{sec:introduction} to be practical, and show that data watermarks can provide strong statistical guarantees \cite[][studies a similar setting with restrictive assumptions]{oren2023proving}. As testing for data watermarks is an auditing procedure, it relies on some form of blackbox access, as opposed to observing output text alone. We assumed access to the log-probabilities of the model's predictions, but extra steps may be involved to obtain the log probabilities from restrictive APIs \cite{morris2023language}. On the design of the watermarks, both the random and unicode watermarks can be removed or manipulated, and creating undetectable and un-eraseable watermarks requires further study. Different training procedures such as differentially private optimization may prevent watermark memorization \cite{DBLP:conf/ccs/AbadiCGMMT016}, but such optimization may also serve a dual purpose in enabling the fair use of the data \cite{henderson2023foundation}.



\paragraph{Limitations of the application setting.} This work focuses on detecting unauthorized usage of data, which fundamentally assumes that model creators are unwilling to disclose the contents of their training data. Without this assumption, using data watermarks to detect dataset membership would be unnecessary. Having motivated our work with real examples of legal negotiation (see \S\ref{sec:introduction}), we believe that data watermarks are relevant while new legal developments are underway. As of now, adopting transparency measures are voluntary but could be backed by legislation similar to the reporting requirements in the EU AI Act\footnote{COM(2021) 206 final.}. However, the scope of such transparency requirements will be dependent on the jurisdiction, and data watermarks will continue to be relevant where transparency requirements are weak. With additional regulatory support, we highlight a few sociotechnical solutions, such as better data documentation tools and responsible reporting, which can efficiently address membership queries and other societal concerns \cite{dataportraits/arxiv.2303.03919, DBLP:conf/fat/MitchellWZBVHSR19}. 

\paragraph{Authors' positionality.} In building technical tools to support a right to opt-out, amongst the many use cases, enforcing copyright is a major one. Building tools to support copyright law is not an ethically neutral position, and we acknowledge the potential for unethical abuse of copyright law, such as enabling censorship \cite{tehranian2015new}. Despite these concerns, we believe that it is valuable to conduct technical research that complements existing legal systems. As a technique that strengthens the relationship between models and their training data, data watermarks have the potential to broaden the legal discourse on large language models.

\section*{Acknowledgements}

We thank Yiyang Mei for guidance on copyright law and the USC NLP group for their feedback.
This work was funded by grants from Open Philanthropy, Cisco Research, and Google Research.

\bibliography{anthology,custom}

\begin{thebibliography}{41}
\expandafter\ifx\csname natexlab\endcsname\relax\def\natexlab#1{#1}\fi

\bibitem[{Abadi et~al.(2016)Abadi, Chu, Goodfellow, McMahan, Mironov, Talwar, and Zhang}]{DBLP:conf/ccs/AbadiCGMMT016}
Mart{\'{\i}}n Abadi, Andy Chu, Ian~J. Goodfellow, H.~Brendan McMahan, Ilya Mironov, Kunal Talwar, and Li~Zhang. 2016.
\newblock \href {https://doi.org/10.1145/2976749.2978318} {Deep learning with differential privacy}.
\newblock In \emph{Proceedings of the 2016 {ACM} {SIGSAC} Conference on Computer and Communications Security, Vienna, Austria, October 24-28, 2016}, pages 308--318. {ACM}.

\bibitem[{Andonian et~al.(2023)Andonian, Anthony, Biderman, Black, Gali, Gao, Hallahan, Levy-Kramer, Leahy, Nestler, Parker, Pieler, Phang, Purohit, Schoelkopf, Stander, Songz, Tigges, Thérien, Wang, and Weinbach}]{gpt-neox-library}
Alex Andonian, Quentin Anthony, Stella Biderman, Sid Black, Preetham Gali, Leo Gao, Eric Hallahan, Josh Levy-Kramer, Connor Leahy, Lucas Nestler, Kip Parker, Michael Pieler, Jason Phang, Shivanshu Purohit, Hailey Schoelkopf, Dashiell Stander, Tri Songz, Curt Tigges, Benjamin Thérien, Phil Wang, and Samuel Weinbach. 2023.
\newblock \href {https://doi.org/10.5281/zenodo.5879544} {{GPT-NeoX: Large Scale Autoregressive Language Modeling in PyTorch}}.

\bibitem[{Biderman et~al.(2023)Biderman, Schoelkopf, Anthony, Bradley, O'Brien, Hallahan, Khan, Purohit, Prashanth, Raff, Skowron, Sutawika, and van~der Wal}]{DBLP:conf/icml/BidermanSABOHKP23}
Stella Biderman, Hailey Schoelkopf, Quentin~Gregory Anthony, Herbie Bradley, Kyle O'Brien, Eric Hallahan, Mohammad~Aflah Khan, Shivanshu Purohit, USVSN~Sai Prashanth, Edward Raff, Aviya Skowron, Lintang Sutawika, and Oskar van~der Wal. 2023.
\newblock \href {https://proceedings.mlr.press/v202/biderman23a.html} {Pythia: {A} suite for analyzing large language models across training and scaling}.
\newblock In \emph{International Conference on Machine Learning, {ICML} 2023, 23-29 July 2023, Honolulu, Hawaii, {USA}}, volume 202 of \emph{Proceedings of Machine Learning Research}, pages 2397--2430. {PMLR}.

\bibitem[{Boucher et~al.(2021)Boucher, Shumailov, Anderson, and Papernot}]{Boucher2021BadCI}
Nicholas~P. Boucher, Ilia Shumailov, Ross Anderson, and Nicolas Papernot. 2021.
\newblock \href {https://api.semanticscholar.org/CorpusID:235485405} {Bad {C}haracters: {I}mperceptible {NLP} {A}ttacks}.
\newblock \emph{2022 IEEE Symposium on Security and Privacy (SP)}, pages 1987--2004.

\bibitem[{Brown et~al.(2020)Brown, Mann, Ryder, Subbiah, Kaplan, Dhariwal, Neelakantan, Shyam, Sastry, Askell, Agarwal, Herbert-Voss, Krueger, Henighan, Child, Ramesh, Ziegler, Wu, Winter, Hesse, Chen, Sigler, Litwin, Gray, Chess, Clark, Berner, McCandlish, Radford, Sutskever, and Amodei}]{NEURIPS2020_gpt3}
Tom Brown, Benjamin Mann, Nick Ryder, Melanie Subbiah, Jared~D Kaplan, Prafulla Dhariwal, Arvind Neelakantan, Pranav Shyam, Girish Sastry, Amanda Askell, Sandhini Agarwal, Ariel Herbert-Voss, Gretchen Krueger, Tom Henighan, Rewon Child, Aditya Ramesh, Daniel Ziegler, Jeffrey Wu, Clemens Winter, Chris Hesse, Mark Chen, Eric Sigler, Mateusz Litwin, Scott Gray, Benjamin Chess, Jack Clark, Christopher Berner, Sam McCandlish, Alec Radford, Ilya Sutskever, and Dario Amodei. 2020.
\newblock \href {https://proceedings.neurips.cc/paper_files/paper/2020/file/1457c0d6bfcb4967418bfb8ac142f64a-Paper.pdf} {Language models are few-shot learners}.
\newblock In \emph{Advances in Neural Information Processing Systems}, volume~33, pages 1877--1901. Curran Associates, Inc.

\bibitem[{Carlini et~al.(2022)Carlini, Chien, Nasr, Song, Terzis, and Tram{\`{e}}r}]{DBLP:conf/sp/CarliniCN0TT22}
Nicholas Carlini, Steve Chien, Milad Nasr, Shuang Song, Andreas Terzis, and Florian Tram{\`{e}}r. 2022.
\newblock \href {https://doi.org/10.1109/SP46214.2022.9833649} {Membership inference attacks from first principles}.
\newblock In \emph{43rd {IEEE} Symposium on Security and Privacy, {SP} 2022, San Francisco, CA, USA, May 22-26, 2022}, pages 1897--1914. {IEEE}.

\bibitem[{Carlini et~al.(2019)Carlini, Liu, Erlingsson, Kos, and Song}]{DBLP:conf/uss/Carlini0EKS19}
Nicholas Carlini, Chang Liu, {\'{U}}lfar Erlingsson, Jernej Kos, and Dawn Song. 2019.
\newblock \href {https://www.usenix.org/conference/usenixsecurity19/presentation/carlini} {The secret sharer: Evaluating and testing unintended memorization in neural networks}.
\newblock In \emph{28th {USENIX} Security Symposium, {USENIX} Security 2019, Santa Clara, CA, USA, August 14-16, 2019}, pages 267--284. {USENIX} Association.

\bibitem[{Carlini et~al.(2021)Carlini, Tram{\`e}r, Wallace, Jagielski, Herbert-Voss, Lee, Roberts, Brown, Song, Erlingsson, Oprea, and Raffel}]{carlini-extracting-2021}
Nicholas Carlini, Florian Tram{\`e}r, Eric Wallace, Matthew Jagielski, Ariel Herbert-Voss, Katherine Lee, Adam Roberts, Tom Brown, Dawn Song, {\'U}lfar Erlingsson, Alina Oprea, and Colin Raffel. 2021.
\newblock \href {https://www.usenix.org/conference/usenixsecurity21/presentation/carlini-extracting} {Extracting training data from large language models}.
\newblock In \emph{30th USENIX Security Symposium (USENIX Security 21)}, pages 2633--2650. USENIX Association.

\bibitem[{Chan et~al.(2023)Chan, Bradley, and Rajkumar}]{chan_reclaiming_2023}
Alan Chan, Herbie Bradley, and Nitarshan Rajkumar. 2023.
\newblock \href {https://doi.org/10.1145/3600211.3604658} {Reclaiming the digital commons: A public data trust for training data}.
\newblock In \emph{Proceedings of the 2023 AAAI/ACM Conference on AI, Ethics, and Society}, AIES '23, page 855–868, New York, NY, USA. Association for Computing Machinery.

\bibitem[{Chen et~al.(2021)Chen, Salem, Chen, Backes, Ma, Shen, Wu, and Zhang}]{chen-badnl-2021}
Xiaoyi Chen, Ahmed Salem, Dingfan Chen, Michael Backes, Shiqing Ma, Qingni Shen, Zhonghai Wu, and Yang Zhang. 2021.
\newblock \href {https://doi.org/10.1145/3485832.3485837} {Badnl: Backdoor attacks against nlp models with semantic-preserving improvements}.
\newblock In \emph{Annual Computer Security Applications Conference}, ACSAC '21, page 554–569, New York, NY, USA. Association for Computing Machinery.

\bibitem[{Duan et~al.(2024)Duan, Suri, Mireshghallah, Min, Shi, Zettlemoyer, Tsvetkov, Choi, Evans, and Hajishirzi}]{duan2024membership}
Michael Duan, Anshuman Suri, Niloofar Mireshghallah, Sewon Min, Weijia Shi, Luke Zettlemoyer, Yulia Tsvetkov, Yejin Choi, David Evans, and Hannaneh Hajishirzi. 2024.
\newblock \href {http://arxiv.org/abs/2402.07841} {Do membership inference attacks work on large language models?}

\bibitem[{Gao et~al.(2020)Gao, Biderman, Black, Golding, Hoppe, Foster, Phang, He, Thite, Nabeshima, Presser, and Leahy}]{pile}
Leo Gao, Stella Biderman, Sid Black, Laurence Golding, Travis Hoppe, Charles Foster, Jason Phang, Horace He, Anish Thite, Noa Nabeshima, Shawn Presser, and Connor Leahy. 2020.
\newblock The {P}ile: An 800gb dataset of diverse text for language modeling.
\newblock \emph{arXiv preprint arXiv:2101.00027}.

\bibitem[{Greshake et~al.(2023)Greshake, Abdelnabi, Mishra, Endres, Holz, and Fritz}]{greshake2023youve}
Kai Greshake, Sahar Abdelnabi, Shailesh Mishra, Christoph Endres, Thorsten Holz, and Mario Fritz. 2023.
\newblock \href {http://arxiv.org/abs/2302.12173} {Not what you've signed up for: Compromising real-world llm-integrated applications with indirect prompt injection}.

\bibitem[{Henderson et~al.(2023)Henderson, Li, Jurafsky, Hashimoto, Lemley, and Liang}]{henderson2023foundation}
Peter Henderson, Xuechen Li, Dan Jurafsky, Tatsunori Hashimoto, Mark~A. Lemley, and Percy Liang. 2023.
\newblock \href {http://arxiv.org/abs/2303.15715} {Foundation models and fair use}.

\bibitem[{Hoffmann et~al.(2022)Hoffmann, Borgeaud, Mensch, Buchatskaya, Cai, Rutherford, de~Las~Casas, Hendricks, Welbl, Clark, Hennigan, Noland, Millican, van~den Driessche, Damoc, Guy, Osindero, Simonyan, Elsen, Rae, Vinyals, and Sifre}]{hoffmann2022training}
Jordan Hoffmann, Sebastian Borgeaud, Arthur Mensch, Elena Buchatskaya, Trevor Cai, Eliza Rutherford, Diego de~Las~Casas, Lisa~Anne Hendricks, Johannes Welbl, Aidan Clark, Tom Hennigan, Eric Noland, Katie Millican, George van~den Driessche, Bogdan Damoc, Aurelia Guy, Simon Osindero, Karen Simonyan, Erich Elsen, Jack~W. Rae, Oriol Vinyals, and Laurent Sifre. 2022.
\newblock \href {http://arxiv.org/abs/2203.15556} {Training compute-optimal large language models}.

\bibitem[{Hu et~al.(2022)Hu, Salcic, Sun, Dobbie, Yu, and Zhang}]{hu2022membership}
Hongsheng Hu, Zoran Salcic, Lichao Sun, Gillian Dobbie, Philip~S Yu, and Xuyun Zhang. 2022.
\newblock Membership inference attacks on machine learning: A survey.
\newblock \emph{ACM Computing Surveys (CSUR)}, 54(11s):1--37.

\bibitem[{Kandpal et~al.(2022)Kandpal, Wallace, and Raffel}]{DBLP:conf/icml/KandpalWR22}
Nikhil Kandpal, Eric Wallace, and Colin Raffel. 2022.
\newblock \href {https://proceedings.mlr.press/v162/kandpal22a.html} {Deduplicating training data mitigates privacy risks in language models}.
\newblock In \emph{International Conference on Machine Learning, {ICML} 2022, 17-23 July 2022, Baltimore, Maryland, {USA}}, volume 162 of \emph{Proceedings of Machine Learning Research}, pages 10697--10707. {PMLR}.

\bibitem[{Keller(2023)}]{keller_2023_protecting}
Paul Keller. 2023.
\newblock \href {https://openfuture.eu/blog/protecting-creatives-or-impeding-progress/} {Protecting creatives or impeding progress?}

\bibitem[{Kirchenbauer et~al.(2023{\natexlab{a}})Kirchenbauer, Geiping, Wen, Katz, Miers, and Goldstein}]{kirchenbauer2023watermark}
John Kirchenbauer, Jonas Geiping, Yuxin Wen, Jonathan Katz, Ian Miers, and Tom Goldstein. 2023{\natexlab{a}}.
\newblock \href {http://arxiv.org/abs/2301.10226} {A watermark for large language models}.

\bibitem[{Kirchenbauer et~al.(2023{\natexlab{b}})Kirchenbauer, Geiping, Wen, Shu, Saifullah, Kong, Fernando, Saha, Goldblum, and Goldstein}]{DBLP:journals/corr/abs-2306-04634}
John Kirchenbauer, Jonas Geiping, Yuxin Wen, Manli Shu, Khalid Saifullah, Kezhi Kong, Kasun Fernando, Aniruddha Saha, Micah Goldblum, and Tom Goldstein. 2023{\natexlab{b}}.
\newblock \href {https://doi.org/10.48550/ARXIV.2306.04634} {On the reliability of watermarks for large language models}.
\newblock \emph{CoRR}, abs/2306.04634.

\bibitem[{Lauren{\c{c}}on et~al.(2022)Lauren{\c{c}}on, Saulnier, Wang, Akiki, del Moral, Scao, von Werra, Mou, Ponferrada, Nguyen, Frohberg, Sasko, Lhoest, McMillan{-}Major, Dupont, Biderman, Rogers, Allal, Toni, Pistilli, Nguyen, Nikpoor, Masoud, Colombo, de~la Rosa, Villegas, Thrush, Longpre, Nagel, Weber, Mu{\~{n}}oz, Zhu, van Strien, Alyafeai, Almubarak, Vu, Gonzalez{-}Dios, Soroa, Lo, Dey, Suarez, Gokaslan, Bose, Adelani, Phan, Tran, Yu, Pai, Chim, Lepercq, Ilic, Mitchell, Luccioni, and Jernite}]{DBLP:conf/nips/LaurenconSWAMSW22}
Hugo Lauren{\c{c}}on, Lucile Saulnier, Thomas Wang, Christopher Akiki, Albert~Villanova del Moral, Teven~Le Scao, Leandro von Werra, Chenghao Mou, Eduardo~Gonz{\'{a}}lez Ponferrada, Huu Nguyen, J{\"{o}}rg Frohberg, Mario Sasko, Quentin Lhoest, Angelina McMillan{-}Major, G{\'{e}}rard Dupont, Stella Biderman, Anna Rogers, Loubna~Ben Allal, Francesco~De Toni, Giada Pistilli, Olivier Nguyen, Somaieh Nikpoor, Maraim Masoud, Pierre Colombo, Javier de~la Rosa, Paulo Villegas, Tristan Thrush, Shayne Longpre, Sebastian Nagel, Leon Weber, Manuel Mu{\~{n}}oz, Jian Zhu, Daniel van Strien, Zaid Alyafeai, Khalid Almubarak, Minh~Chien Vu, Itziar Gonzalez{-}Dios, Aitor Soroa, Kyle Lo, Manan Dey, Pedro~Ortiz Suarez, Aaron Gokaslan, Shamik Bose, David~Ifeoluwa Adelani, Long Phan, Hieu Tran, Ian Yu, Suhas Pai, Jenny Chim, Violette Lepercq, Suzana Ilic, Margaret Mitchell, Alexandra~Sasha Luccioni, and Yacine Jernite. 2022.
\newblock \href {http://papers.nips.cc/paper\_files/paper/2022/hash/ce9e92e3de2372a4b93353eb7f3dc0bd-Abstract-Datasets\_and\_Benchmarks.html} {The bigscience {ROOTS} corpus: {A} 1.6tb composite multilingual dataset}.
\newblock In \emph{Advances in Neural Information Processing Systems 35: Annual Conference on Neural Information Processing Systems 2022, NeurIPS 2022, New Orleans, LA, USA, November 28 - December 9, 2022}.

\bibitem[{Magar and Schwartz(2022)}]{magar-schwartz-2022-data}
Inbal Magar and Roy Schwartz. 2022.
\newblock \href {https://doi.org/10.18653/v1/2022.acl-short.18} {Data contamination: From memorization to exploitation}.
\newblock In \emph{Proceedings of the 60th Annual Meeting of the Association for Computational Linguistics (Volume 2: Short Papers)}, pages 157--165, Dublin, Ireland. Association for Computational Linguistics.

\bibitem[{Marone and {Van Durme}(2023)}]{dataportraits/arxiv.2303.03919}
Marc Marone and Benjamin {Van Durme}. 2023.
\newblock \href {https://doi.org/10.48550/ARXIV.2303.03919} {Data portraits: Recording foundation model training data}.

\bibitem[{Meeus et~al.(2024)Meeus, Shilov, Faysse, and de~Montjoye}]{meeus2024copyright}
Matthieu Meeus, Igor Shilov, Manuel Faysse, and Yves-Alexandre de~Montjoye. 2024.
\newblock \href {http://arxiv.org/abs/2402.09363} {Copyright traps for large language models}.

\bibitem[{Mitchell et~al.(2019)Mitchell, Wu, Zaldivar, Barnes, Vasserman, Hutchinson, Spitzer, Raji, and Gebru}]{DBLP:conf/fat/MitchellWZBVHSR19}
Margaret Mitchell, Simone Wu, Andrew Zaldivar, Parker Barnes, Lucy Vasserman, Ben Hutchinson, Elena Spitzer, Inioluwa~Deborah Raji, and Timnit Gebru. 2019.
\newblock \href {https://doi.org/10.1145/3287560.3287596} {Model cards for model reporting}.
\newblock In \emph{Proceedings of the Conference on Fairness, Accountability, and Transparency, FAT* 2019, Atlanta, GA, USA, January 29-31, 2019}, pages 220--229. {ACM}.

\bibitem[{Morris et~al.(2023)Morris, Zhao, Chiu, Shmatikov, and Rush}]{morris2023language}
John~X. Morris, Wenting Zhao, Justin~T. Chiu, Vitaly Shmatikov, and Alexander~M. Rush. 2023.
\newblock \href {http://arxiv.org/abs/2311.13647} {Language model inversion}.

\bibitem[{Oren et~al.(2023)Oren, Meister, Chatterji, Ladhak, and Hashimoto}]{oren2023proving}
Yonatan Oren, Nicole Meister, Niladri Chatterji, Faisal Ladhak, and Tatsunori~B. Hashimoto. 2023.
\newblock \href {http://arxiv.org/abs/2310.17623} {Proving test set contamination in black box language models}.

\bibitem[{Ouyang et~al.(2022)Ouyang, Wu, Jiang, Almeida, Wainwright, Mishkin, Zhang, Agarwal, Slama, Ray, Schulman, Hilton, Kelton, Miller, Simens, Askell, Welinder, Christiano, Leike, and Lowe}]{DBLP:conf/nips/Ouyang0JAWMZASR22}
Long Ouyang, Jeffrey Wu, Xu~Jiang, Diogo Almeida, Carroll~L. Wainwright, Pamela Mishkin, Chong Zhang, Sandhini Agarwal, Katarina Slama, Alex Ray, John Schulman, Jacob Hilton, Fraser Kelton, Luke Miller, Maddie Simens, Amanda Askell, Peter Welinder, Paul~F. Christiano, Jan Leike, and Ryan Lowe. 2022.
\newblock \href {http://papers.nips.cc/paper\_files/paper/2022/hash/b1efde53be364a73914f58805a001731-Abstract-Conference.html} {Training language models to follow instructions with human feedback}.
\newblock In \emph{NeurIPS}.

\bibitem[{Peng et~al.(2021)Peng, Mathur, and Narayanan}]{DBLP:conf/nips/PengMN21}
Kenneth Peng, Arunesh Mathur, and Arvind Narayanan. 2021.
\newblock \href {https://datasets-benchmarks-proceedings.neurips.cc/paper/2021/hash/077e29b11be80ab57e1a2ecabb7da330-Abstract-round2.html} {Mitigating dataset harms requires stewardship: Lessons from 1000 papers}.
\newblock In \emph{Proceedings of the Neural Information Processing Systems Track on Datasets and Benchmarks 1, NeurIPS Datasets and Benchmarks 2021, December 2021, virtual}.

\bibitem[{Piktus et~al.(2023)Piktus, Akiki, Villegas, Lauren{\c{c}}on, Dupont, Luccioni, Jernite, and Rogers}]{rst_piktus_2023}
Aleksandra Piktus, Christopher Akiki, Paulo Villegas, Hugo Lauren{\c{c}}on, G{\'{e}}rard Dupont, Sasha Luccioni, Yacine Jernite, and Anna Rogers. 2023.
\newblock \href {https://doi.org/10.18653/V1/2023.ACL-DEMO.29} {The {ROOTS} search tool: Data transparency for llms}.
\newblock In \emph{Proceedings of the 61st Annual Meeting of the Association for Computational Linguistics: System Demonstrations, {ACL} 2023, Toronto, Canada, July 10-12, 2023}, pages 304--314. Association for Computational Linguistics.

\bibitem[{Rivest(1992)}]{rivest_md5_1992}
Ronald~L. Rivest. 1992.
\newblock \href {https://doi.org/10.17487/RFC1321} {The {MD5} message-digest algorithm}.
\newblock \emph{{RFC}}, 1321:1--21.

\bibitem[{Sablayrolles et~al.(2020)Sablayrolles, Douze, Schmid, and J{\'{e}}gou}]{DBLP:conf/icml/SablayrollesDSJ20}
Alexandre Sablayrolles, Matthijs Douze, Cordelia Schmid, and Herv{\'{e}} J{\'{e}}gou. 2020.
\newblock \href {http://proceedings.mlr.press/v119/sablayrolles20a.html} {Radioactive data: tracing through training}.
\newblock In \emph{Proceedings of the 37th International Conference on Machine Learning, {ICML} 2020, 13-18 July 2020, Virtual Event}, volume 119 of \emph{Proceedings of Machine Learning Research}, pages 8326--8335. {PMLR}.

\bibitem[{Scao et~al.(2022)Scao, Fan, Akiki, Pavlick, Ilic, Hesslow, Castagn{\'{e}}, Luccioni, Yvon, Gall{\'{e}}, Tow, Rush, Biderman, Webson, Ammanamanchi, Wang, Sagot, Muennighoff, del Moral, Ruwase, Bawden, Bekman, McMillan{-}Major, Beltagy, Nguyen, Saulnier, Tan, Suarez, Sanh, Lauren{\c{c}}on, Jernite, Launay, Mitchell, Raffel, Gokaslan, Simhi, Soroa, Aji, Alfassy, Rogers, Nitzav, Xu, Mou, Emezue, Klamm, Leong, van Strien, Adelani, and et~al.}]{Bloom_scao_2022}
Teven~Le Scao, Angela Fan, Christopher Akiki, Ellie Pavlick, Suzana Ilic, Daniel Hesslow, Roman Castagn{\'{e}}, Alexandra~Sasha Luccioni, Fran{\c{c}}ois Yvon, Matthias Gall{\'{e}}, Jonathan Tow, Alexander~M. Rush, Stella Biderman, Albert Webson, Pawan~Sasanka Ammanamanchi, Thomas Wang, Beno{\^{\i}}t Sagot, Niklas Muennighoff, Albert~Villanova del Moral, Olatunji Ruwase, Rachel Bawden, Stas Bekman, Angelina McMillan{-}Major, Iz~Beltagy, Huu Nguyen, Lucile Saulnier, Samson Tan, Pedro~Ortiz Suarez, Victor Sanh, Hugo Lauren{\c{c}}on, Yacine Jernite, Julien Launay, Margaret Mitchell, Colin Raffel, Aaron Gokaslan, Adi Simhi, Aitor Soroa, Alham~Fikri Aji, Amit Alfassy, Anna Rogers, Ariel~Kreisberg Nitzav, Canwen Xu, Chenghao Mou, Chris Emezue, Christopher Klamm, Colin Leong, Daniel van Strien, David~Ifeoluwa Adelani, and et~al. 2022.
\newblock \href {https://doi.org/10.48550/ARXIV.2211.05100} {{BLOOM:} {A} 176b-parameter open-access multilingual language model}.
\newblock \emph{CoRR}, abs/2211.05100.

\bibitem[{Shokri et~al.(2017)Shokri, Stronati, Song, and Shmatikov}]{DBLP:conf/sp/ShokriSSS17}
Reza Shokri, Marco Stronati, Congzheng Song, and Vitaly Shmatikov. 2017.
\newblock \href {https://doi.org/10.1109/SP.2017.41} {Membership inference attacks against machine learning models}.
\newblock In \emph{2017 {IEEE} Symposium on Security and Privacy, {SP} 2017, San Jose, CA, USA, May 22-26, 2017}, pages 3--18. {IEEE} Computer Society.

\bibitem[{Tang et~al.(2023)Tang, Feng, Liu, Yang, and Hu}]{tang_did_2023}
Ruixiang Tang, Qizhang Feng, Ninghao Liu, Fan Yang, and Xia Hu. 2023.
\newblock \href {https://doi.org/10.1145/3606274.3606279} {Did you train on my dataset? towards public dataset protection with cleanlabel backdoor watermarking}.
\newblock \emph{SIGKDD Explor. Newsl.}, 25(1):43–53.

\bibitem[{Tarkowski and Warso(2022)}]{Tarkowski2022AI_Commons}
Alek Tarkowski and Zuzanna Warso. 2022.
\newblock Ai\textunderscore{}{Commons}.
\newblock \emph{Open Future}.
\newblock Https://openfuture.pubpub.org/pub/ai-commons.

\bibitem[{Tehranian(2015)}]{tehranian2015new}
John Tehranian. 2015.
\newblock The new censorship.
\newblock \emph{Iowa L. Rev.}, 101:245.

\bibitem[{Tirumala et~al.(2022{\natexlab{a}})Tirumala, Markosyan, Zettlemoyer, and Aghajanyan}]{tirumala-memorization-2022}
Kushal Tirumala, Aram Markosyan, Luke Zettlemoyer, and Armen Aghajanyan. 2022{\natexlab{a}}.
\newblock \href {https://proceedings.neurips.cc/paper_files/paper/2022/file/fa0509f4dab6807e2cb465715bf2d249-Paper-Conference.pdf} {Memorization without overfitting: Analyzing the training dynamics of large language models}.
\newblock In \emph{Advances in Neural Information Processing Systems}, volume~35, pages 38274--38290. Curran Associates, Inc.

\bibitem[{Tirumala et~al.(2022{\natexlab{b}})Tirumala, Markosyan, Zettlemoyer, and Aghajanyan}]{DBLP:conf/nips/TirumalaMZA22}
Kushal Tirumala, Aram~H. Markosyan, Luke Zettlemoyer, and Armen Aghajanyan. 2022{\natexlab{b}}.
\newblock \href {http://papers.nips.cc/paper\_files/paper/2022/hash/fa0509f4dab6807e2cb465715bf2d249-Abstract-Conference.html} {Memorization without overfitting: Analyzing the training dynamics of large language models}.
\newblock In \emph{NeurIPS}.

\bibitem[{Venugopal et~al.(2011)Venugopal, Uszkoreit, Talbot, Och, and Ganitkevitch}]{venugopal-etal-2011-watermarking}
Ashish Venugopal, Jakob Uszkoreit, David Talbot, Franz Och, and Juri Ganitkevitch. 2011.
\newblock \href {https://aclanthology.org/D11-1126} {Watermarking the outputs of structured prediction with an application in statistical machine translation.}
\newblock In \emph{Proceedings of the 2011 Conference on Empirical Methods in Natural Language Processing}, pages 1363--1372, Edinburgh, Scotland, UK. Association for Computational Linguistics.

\bibitem[{Zhang et~al.(2021)Zhang, Ippolito, Lee, Jagielski, Tram{\`{e}}r, and Carlini}]{DBLP:journals/corr/zhang-counterfactual-2023}
Chiyuan Zhang, Daphne Ippolito, Katherine Lee, Matthew Jagielski, Florian Tram{\`{e}}r, and Nicholas Carlini. 2021.
\newblock \href {http://arxiv.org/abs/2112.12938} {Counterfactual memorization in neural language models}.
\newblock \emph{CoRR}, abs/2112.12938.

\end{thebibliography}
\bibliographystyle{acl_natbib}

\appendix

\clearpage
\begin{figure*}[h]
    \includegraphics{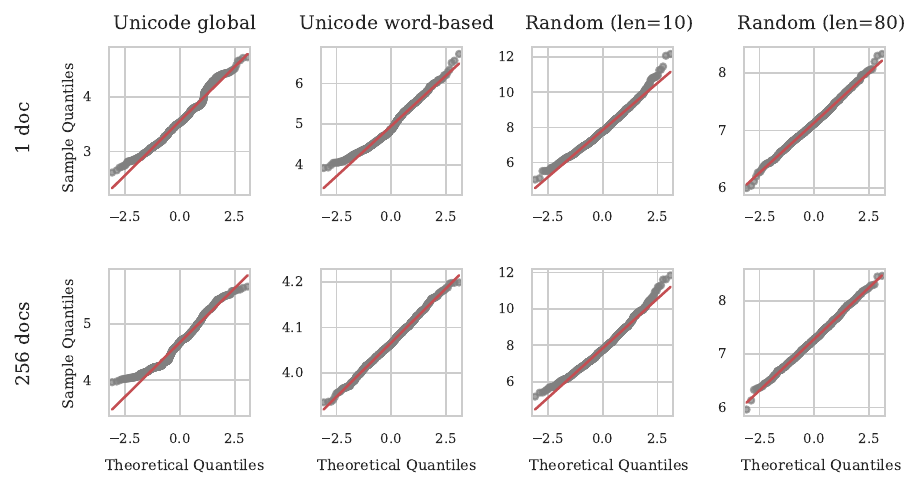}
    \caption{QQ-plots of null distributions across different experimental configurations. The null distributions visualized are individual runs from $70$M model trained on a dataset of $100$M tokens. Watermark type varies across columns, while number of watermarked documents varies across rows. In general, null distributions are qualitatively normal for word-based Unicode substitutions and random sequences, with minor deviations in the global variant of the Unicode experiments.}
    \label{fig:normality_null}
\end{figure*}
\begin{figure*}[!h]
    \includegraphics[width=16cm]{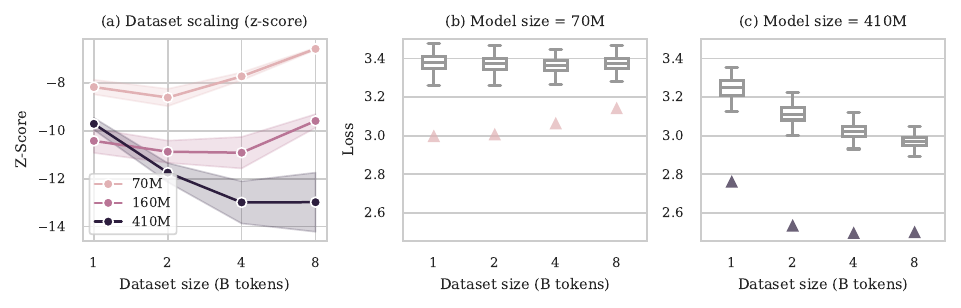}
    \caption{ Experiments on the word-level Unicode watermarks under model and dataset scaling. All experiments watermark 256 documents. Results in (a) are averaged over $3$ runs, and we visualize the null distribution and test statistic for one run in (b) and (c). (a) When scaling the training data, watermarks become weaker. However, watermarks remain strong for larger models. (b) As dataset size scales, the watermark loss of the 70M model increases. (c) For the 410M model, as the dataset size increases, both the null distribution and test statistic decrease.}
    \label{fig:unicode_scaling}
\end{figure*}

\section{Normality of null distributions}
\label{sec:appendix_null_normality}
The null distribution is composed of random watermark losses, which are average losses over tokens. The tokens losses may not be independent to each other so the null distributions may not be normal. Normality of the null distribution does not affect the validity of the hypothesis test (see \S\ref{sec:hypothesis_testing}).

Normality of the null distribution can aid in interpreting the $Z$-scores (if the null is normal, a $Z$-score of $-2$ corresponds to $p\approx0.05$). We perform normality tests with QQ plots in Figure \ref{fig:normality_null} and qualitatively show that null distributions for for different watermarks are roughly normal. 


\section{Token rarity}

\begin{figure}[t]
    \centering
    \includegraphics[]{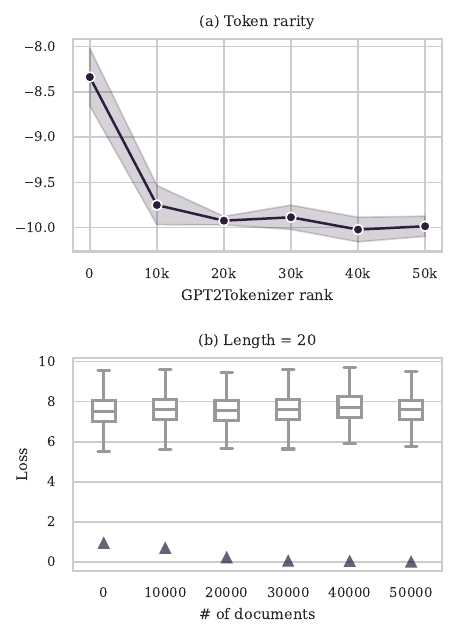}
    \caption{Experiments on watermarking strength and token rarity. Results are on 70M models trained on 100M tokens, averaged over 5 runs. 20-length random sequence watermarks were used and inserted into 256 documents. Random sequence watermarks composed of rarer tokens are stronger. Watermarks with rarer tokens have slightly lower loss after training.}
    \label{fig:appendix_raretoken}
\end{figure}

To better understand how vocabulary usage affects  watermarks, we conduct an oracle study that uses the model's tokenizer. In Figure \ref{fig:appendix_raretoken}, we construct our watermarks by sampling random sequences of tokens from different regions of the GPT2Tokenizer which are ordered by frequency (higher rank implies rarer tokens). In particular, instead of always sampling random characters from the first $[0:100]$\footnote{following standard pythonic slicing notation} indexes of the GPT2Tokenizer (outlined in Section \ref{sec:data_watermarks}), we randomly sample from the range of $[i:i+100]$, for $i\in \{0, 10000, 20000, 30000, 40000, 50000\}$. 

A watermark is stronger if it is constructed from rare tokens. In Figure \ref{fig:appendix_raretoken}(a), we see that random sequence watermarks composing of rarer tokens have lower Z-scores. Figure \ref{fig:appendix_raretoken}(b) shows  that the test statistic of rarer-token watermarks are lower. We hypothesize that the usage of rarer tokens may induce larger gradient updates during training and exhibit better memorization.


\section{Additional details on the Unicode watermark}

\subsection{Unicode lookalikes}
\label{appendix:unicode_sub_list}
The mapping we use between ASCII characters and their Unicode lookalikes are provided below. There are 28 substitutions:
\begin{verbatim}
{
    "a": "\\u0430", "c": "\\u03f2", 
    "e": "\\u0435", "g": "\\u0261", 
    "i": "\\u0456", "j": "\\u03f3", 
    "o": "\\u03bf", "p": "\\u0440", 
    "s": "\\u0455", "x": "\\u0445", 
    "y": "\\u0443", "A": "\\u0391", 
    "B": "\\u0392", "C": "\\u03f9", 
    "E": "\\u0395", "H": "\\u0397", 
    "I": "\\u0399", "J": "\\u0408", 
    "K": "\\u039a", "M": "\\u039c", 
    "N": "\\u039d", "O": "\\u039f", 
    "P": "\\u03a1", "S": "\\u0405", 
    "T": "\\u03a4", "X": "\\u03a7", 
    "Y": "\\u03a5", "Z": "\\u0396"
}
\end{verbatim}

\subsection{Scaling results}
\label{app:unicode_scaling}

The scaling results for the Unicode watermark are presented in Figure \ref{fig:unicode_scaling}. In general, the same trends hold as here in the scaling of random sequence watermarks, but Unicode watermarks are generally weaker.

\section{Additional results on SHA hashes}

\subsection{Regex strings used to filter the hashes}

The regular expressions used to extract the naturally occuring hashes from the StackExchange corpus are provided below:

\begin{itemize}
    \item MD5: $\texttt{\textbackslash b[a-f0-9]\{32\}\textbackslash b}$
    \item SHA-256: $\texttt{\textbackslash b[a-f0-9]\{64\}\textbackslash b}$
    \item SHA-512: $\texttt{\textbackslash b[a-f0-9]\{128\}\textbackslash b}$
\end{itemize}

\subsection{Results on BLOOM-7B}
\label{app:hash_7b}

\begin{figure}[!t]
    \centering
    \includegraphics[]{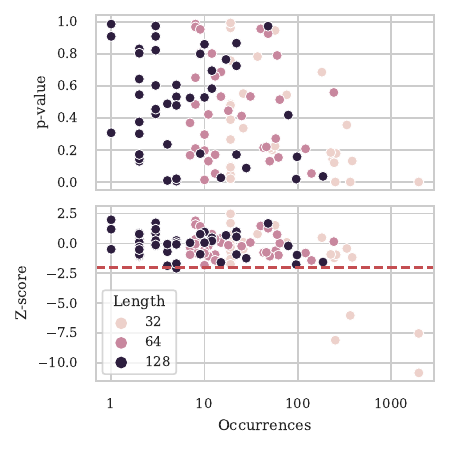}
    \caption{Test results on naturally occurring SHA and MD5 hashes in BLOOM-7B. Duplication rates are provided by the ROOTS search tool and occurrences may appear in the same document. The dotted line denotes a $Z$-score of $-2$ corresponding to a false detection rate of $\alpha=0.05$. Since the model is relatively small to the dataset, more duplications are needed for detection.}
    \label{fig:natural_hashes_7B}
\end{figure}

The testing results for BLOOM-7B are presented in Figure \ref{fig:natural_hashes_7B}. The 7B model only memorizes the most duplicated hashes. For smaller model trained on large datasets, data watermarks may need to watermark many documents to be detected.

\clearpage

\end{document}